\documentclass[
aps,prd,
12pt,
nopreprintnumbers,
showpacs,
eqsecnum,
nofootinbib
]{revtex4-1}

\usepackage{graphicx}

\begin{document}

\title{Hosotani mechanism in higher dimensional Lee-Wick theory}
\author{Nahomi Kan}\email[]{kan@yamaguchi-jc.ac.jp}
\affiliation{
Yamaguchi Junior College,
Hofu-shi, Yamaguchi 747--1232, Japan}
\author{Koichiro Kobayashi}\email[]{m004wa@yamaguchi-u.ac.jp}
\author{Kiyoshi Shiraishi}\email[]{shiraish@yamaguchi-u.ac.jp}
\affiliation{
Yamaguchi University,
Yamaguchi-shi, Yamaguchi 753--8512, Japan}
\date{\today}
\begin{abstract}
Hosotani mechanism in higher-dimensional Lee-Wick theory is
investigated. The symmetry breaking mechanism proposed by Hosotani is
studied at one-loop level through a toy model in this theory. We find
that the phase diagram of symmetry and masses of fields are modified
from the original one if masses of Lee-Wick particles are in the same
order of the inverse of the compactification scale. 
\end{abstract}



\maketitle


\section{Introduction}
Lee-Wick theory has been proposed as a candidate of finite theory for
QED \cite{LW}.  In the theory, a massive ghost field is introduced and
is expected to work as in the manner of Pauli-Villars regularization
scheme. Although the existence of such a ghost field seems to be
problematic for some authors \cite{prob}, possible interpretations have
been suggested, for example, instability of the massive particle
\cite{LW,GOW}.

In spite of of the suspicion on its consistency,
Lee-Wick theory has been applied to not only the standard model
and unification \cite{GOW,LWSM} but also to cosmology of the  bouncing
universe \cite{bounce}. In these days, various physical consequences
have been studied eagerly, even in still higher derivative models
\cite{hd}, while some open questions in their basic concept is left
unresolved. 

In particle physics, it is interesting to apply 
Lee-Wick theory to solve the hierarchy problem.
If the mass of the ghost particles are larger than  TeV scale,
some quadratic divergences and other relic effects become invisible.
Such a model does not need supersymmetry, and we are ready to
study the scenario  \`a la Lee-Wick in the analysis of high-energy
experiments.

In Lee-Wick unified models, however, the origin of the Higgs boson and/or
the structure of symmetry breaking sector are still unexplained. In this
reason, we come to consider so-called gauge-Higgs unification, or the
higher-dimensional origin of the Higgs field \cite{Hosotani,Toms,GH} in
the context of Lee-Wick theory. In the present paper, a simple toy model
with an $SU(2)$ gauge symmetry is examined. In the model, we take a
circle as an extra dimension. For a phenomenologically viable models, we
should use orbifolds for extra dimensions to obtain chiral fermions and
consider selected boundary conditions on gauge and fermionic fields
\cite{GH}. Nevertheless, we will show some remarkable features about
symmetry breaking, which may be worth applying to the phenomenological
study in future work.

Another reason to examine the gauge-Higgs unified scheme is in the
interest on the quantum effect. The tree-level mass of the scalar degree
of freedom, which comes from an extra component of the gauge field, is
protected from quantum corrections because of the gauge symmetry in
the higher-dimensional theory. Then, the quantum effective potential
governs the symmetry breakdown. Whereas in Lee-Wick theory,
the one-loop contribution of the ghost fields modifies the effective
potential in the theory as well as its divergent behavior. We expect new
varieties of symmetry breaking patterns in Lee-Wick gauge-Higgs
unified models.

In this paper, we consider the higher dimensional model of
Lee-Wick theory and the one-loop effect on the symmetry breaking
mechanism proposed by Hosotani. In Sec.~\ref{S2}, we review the Lee-Wick
gauge theory and its extension to higher dimensional theory. 
Then, we incorporate Hosotani mechanism into the five dimensional
Lee-Wick theory and investigate mass levels of the particles
obtained by compactification on $R^4\times S^1$. In Sec.~\ref{S3}, we
calculate the one-loop effective potential and study the symmetry
breaking in our model. We discuss the mass of the vector and scalar
bosons in our model in Sec.~\ref{S4}.  Finally, Sec.~\ref{S5} is devoted
to discussion.

\section{The higher derivative gauge theory in five dimensions
\label{S2}}

The Lagrangian density for the Lee-Wick non-Abelian gauge theory is given
by
\cite{GOW}
\begin{equation}
{\cal L}=-\frac{1}{2}{\rm
tr~}\hat{F}^{\mu\nu}\hat{F}_{\mu\nu}+\frac{1}{M_A^2} {\rm
tr~}\hat{D}_\mu
\hat{F}^{\mu\nu}\hat{D}^{\lambda}\hat{F}_{\lambda\nu}\,,
\label{LWL1}
\end{equation}
where
$\hat{F}_{\mu\nu}\equiv
\partial_\mu\hat{A}_\nu-\partial_\nu\hat{A}_\mu-ig[\hat{A}_\mu ,
\hat{A}_\nu]$ and $g$ is the gauge coupling. By using auxiliary field
$\tilde{A}_\nu$, the Lagrangian (\ref{LWL1}) is rewritten as
\begin{equation}
{\cal L}=-\frac{1}{2}{\rm
tr~}\hat{F}^{\mu\nu}\hat{F}_{\mu\nu}-2 {\rm
tr~}\tilde{A}_\nu\hat{D}_\mu
\hat{F}^{\mu\nu}-{M_A^2}\,{\rm
tr~}\tilde{A}^\nu\tilde{A}_\nu\,.
\end{equation}
In this Lagrangian, the kinetic term of the fields does not
take a diagonal form. If we define a linear combination of the 
fields
$A_\mu\equiv\hat{A}_\mu-\tilde{A}_\mu$, we obtain the
diagonalized Lagrangian \cite{GOW}
\begin{eqnarray}
{\cal L}&=&-\frac{1}{2}{\rm
tr~}{F}^{\mu\nu}{F}_{\mu\nu}+\frac{1}{2}{\rm
tr~}(D^\mu\tilde{A}^\nu-D^\nu\tilde{A}^\mu)
(D_\mu\tilde{A}_\nu-D_\nu\tilde{A}_\mu)
-{M_A^2}\,{\rm
tr~}\tilde{A}^\nu\tilde{A}_\nu\nonumber \\
& &-ig{\rm tr~}([\tilde{A}_\mu, \tilde{A}_\mu] F^{\mu\nu})+\cdots\,,
\end{eqnarray}
where ${F}_{\mu\nu}\equiv
\partial_\mu{A}_\nu-\partial_\nu{A}_\mu-ig[{A}_\mu ,
{A}_\nu]$ and $D_\mu\tilde{A}_\nu\equiv\partial_\mu\tilde{A}_\nu
-ig[{A}_\mu ,\tilde{A}_\nu]$.
Here, we omitted the irrelevant interaction terms for our purpose.
The Lagrangian shows that the theory describes usual massless gauge
bosons and  vector `ghost' bosons with mass $M_A$.

The Lee-Wick gauge theory in higher dimensions can be defined by the same
Lagrangian density. Let us assume the five dimensional theory.
Suppose that the spacetime topology is
$R^4 \times S^1$ where $R^4$ is a four dimensional
spacetime and $S^1$ is a circle whose circumference is $L$. The five
dimensional coordinates are represented by $(x^i, y)=(x^0,x^1,x^2,x^3,
y)$, where
$0\le y<L$. Now, the fifth component of the gauge field $A_5$  becomes
scalar degrees of freedom in four dimensions. 

Let us consider $SU(2)$ gauge theory in this spacetime, as a simple
case. Then, the matrix-valued gauge field can be expressed by three
component fields as $A_\mu=A_\mu^a\frac{\tau^a}{2}$, where $\tau^a$ is
the $2\times 2$ Pauli matrix and $a=1, 2, 3$. Boundary conditions on
$A_\mu$ is chosen definitely as follows: 
\begin{equation}
A_\mu(x, y + L)= A_\mu(x, y)\,.
\end{equation}

A constant vacuum gauge field is allowed on the
multiply-connected space; the vacuum expectation value of the gauge
field configuration plays the role of an `order parameter'. 
Hosotani and Toms have considered a simple model to show that one-loop
vacuum effect determines the order parameter \cite{Hosotani,Toms}. 

On the circle $S^1$ as an extra dimension, non zero vacuum gauge
configuration is permitted. In our case, we set
\begin{equation}
gLA_5=\frac{1}{2}\left[
\begin{array}{cc}\phi & 0\\ 0 & -\phi\end{array}\right]
=\frac{\phi}{2}\tau_3\,,
\end{equation}
where the vacuum gauge field has been diagonalized by using the freedom
of gauge transformations.
Therefore, roughly speaking, the extra-dimensional component of the
gauge field behaves as the Higgs field in the adjoint representation.
We should notice that  
$gLA_5$ is equivalent to  $gLA_5+2\pi
\ell \tau_3$ ($\ell$: integer) under non-singular gauge
transformations. Thus, we can only restrict
$0\le \phi < 4\pi$ by gauge transformations.

In the five dimensional Lee-Wick gauge theory, in view of the
four dimensional flat space, we find that the propagator has ladders
of poles at
\begin{equation}
p^2=\frac{1}{L^2}(2\pi \ell+ \phi)^2\equiv m_1^2(\ell)\equiv
m_2^2(\ell)\qquad (\ell{\rm :~integer})
\end{equation}
which are for the four-dimensional descendant fields of
$A^1_i$ and
$A^2_i$, where $i=0, 1, 2, 3$,
\begin{equation}
p^2=\frac{1}{L^2}(2\pi \ell)^2\equiv m_3^2(\ell)\qquad (\ell{\rm
:~integer})
\end{equation}
which are for the descendant fields of $A^3_i$.
Here, $p^2$ denotes the four dimensional momentum squared, of course.
Note that the fields $A_5^1$ and $A_5^2$ are absorbed by $A_i^1$ and
$A_i^2$ and become the longitudinal components of the massive vector
fields in general.

On the other hand, the Lee-Wick ghost field has poles at
\begin{equation}
p^2=\frac{1}{L^2}(2\pi \ell+
\phi)^2+M_A^2\equiv\tilde{m}_1^2(\ell)\equiv\tilde{m}_2^2(\ell)\qquad
(\ell{\rm :~integer})
\end{equation}
which are for the four-dimensional descendant fields of
$\tilde{A}^1_\mu$ and
$\tilde{A}^2_\mu$,
\begin{equation}
p^2=\frac{1}{L^2}(2\pi \ell)^2+M^2_A\equiv\tilde{m}_3^2(\ell)\qquad
(\ell{\rm :~integer})
\end{equation}
which are for the descendant fields of $\tilde{A}^3_\mu$.
Note that $\tilde{m}^2_a=m^2_a+M_A^2$. In the view of four dimensional
spacetime, $\tilde{A}_\mu$ is decomposed to massive vector and
scalar fields.

In general, non zero $\phi$ shifts the mass level; this effect reduces
the number of the massless fields in four dimensions within the
Kaluza-Klein point of view. We will see how the value of $\phi$ is
determined by the one-loop quantum effect in the next section.

\section{One-Loop Quantum Effect and symmetry breaking
\label{S3}}

To calculate the one-loop effective potential, we use integration of the
Euclidean momenta. The choice of calculation is subtle in the
higher-derivative theory. According to Ref.~\cite{AG}, the Euclidean
approach naturally leads to the cancellation of the leading divergence in
the loop effect, as is suitable for the motivation of Lee-Wick theory.

The one-loop vacuum energy density from the $SU(2)$ gauge and Lee-Wick
fields discussed in the previous section is given by
\begin{equation}
V_{A}=\sum_{a=1}^3\sum_{\ell=-\infty}^\infty
\int\frac{d^4k_E}{(2\pi)^4}\left\{\frac{3}{2}\ln[k_E^2+m_a^2(\ell)]-
2\ln[k_E^2+\tilde{m}_a^2(\ell)]\right\}\,.
\end{equation}
Incidentally, the relative sign appearing in this expression agrees
with the result for thermal free energy considered in
Ref.~\cite{thermo}, provided that we use the Matsubara formalism.

Here, we shall briefly show the calculation of the
potential treated in the text. In the space $R^d\times S^1$, the
effective potential 
is formally expressed by the Schwinger proper-time integral
\cite{Schwinger}.
Thus, we find
\begin{equation}
V_A=-\frac{1}{2(4\pi)^2}\int^\infty_{\Lambda^{-2}}
\frac{dt}{t^3}\, K(t)(3-4e^{-M_A^2t})\,,
\end{equation}
where the kernel function is defined as
\begin{equation}
K(t)=\sum_a\sum_\ell \exp[-m^2_a(\ell)t]\,,
\end{equation}
and $\Lambda$ is the UV cut-off scale.

To evaluate the kernel function $K(t)$, we can use the inversion
relation
\begin{equation}
\sum_{\ell=-\infty}^\infty\exp
\left[-\left(\frac{2\pi}{L}\right)^2(\ell-\chi)^2\,t\right]=
\frac{L}{\sqrt{4\pi\, t}}\sum_{\ell=-\infty}^\infty\exp
\left(-\frac{L^2\ell^2}{4t}\right)\cos(2\pi\ell\chi)\,.
\end{equation}
Using this formula, we can see that the UV divergent part comes from
the contribution of $\ell=0$ in the sum on the right-hand side of the
expression and it is independent of $\phi$. Thus, we obtain a
finite one-loop effective potential for $\Lambda\rightarrow\infty$ by
discarding the divergent constant. Then, the integration over the
proper-time can be carried out by the help of an integral representation
of the modified Bessel function of the second kind
\cite{Takenaga}
\begin{equation}
\int_0^\infty\frac{dt}{t^{\nu+1}}\exp\left(-\frac{L^2\ell^2}{4t}-M^2t\right)=
2\left(\frac{2M}{L\ell}\right)^\nu K_\nu(ML\ell)\,.
\end{equation}
Further, an explicit form of $K_{5/2}(z)$
\begin{equation}
K_{5/2}(z)=\sqrt{\frac{\pi}{2z}}\left(1+\frac{3}{z}+\frac{3}{z^2}\right)e^{-z}\,
\end{equation}
leads to the following expression:
\begin{equation}
V_{A}=-\frac{9}{4\pi^2L^4}\sum_{\ell=1}^\infty
\frac{w_A(M_AL\ell)}{\ell^5} [2\cos(\ell\phi)+1]\,,
\end{equation}
where the weight factor $w_A(x)$ is defined as
\begin{equation}
w_A(x)=1-\frac{4}{3}e^{-x}\left(1+x+\frac{1}{3}x^2\right)\,.
\end{equation}

We can see that $w_A(x)$ becomes negative for $x<1.41$.
Thus, the shape of the effective potential can become `reversed' if
$M_AL$ takes a small value, and then the symmetry breaking such that
$\langle\phi\rangle\ne 0$ occurs. 
Unfortunately, because the study on a Lee-Wick model in four dimensions
\cite{GOW} suggests
$M_A>1$TeV, the mass of the vector bosons should be very large in this
case.

In order to naturally realize the symmetry
breaking vacuum, we can add fermionic fields into the theory.
Now, the Lee-Wick Dirac fermion is introduced, which is governed by the
Lagrangian
\begin{equation}
{\cal
L}_{\psi}=-i\bar{\psi}^a
{D\!\!\!\!/}\left(1+\frac{{D\!\!\!\!/}{D\!\!\!\!/}}{M_{\psi}^2}\right)
\psi^a\,,
\end{equation}
where ${D\!\!\!\!/}=\gamma^\mu D_{\mu}$ \cite{GOW,HK}.
If the spacetime is flat with an infinite extension and gauge fields
vanish, the pole of the propagator of the Dirac field is located at
$p^2=0$ and
$p^2=M_{\psi}^2$, and the latter is a ghost pole.

As for the case of the Kaluza-Klein background spacetime, i.e., 
where an extra dimension is compactified on $S^1$ with the radius
$L/(2\pi)$, we obtain the ladder of poles for four dimensional fields.
We consider the $SU(2)$ adjoint fermion $\psi^a$ $(a=1,2,3)$.
Further, we assume a periodic boundary condition in the extra dimension
as
\begin{equation}
\psi^a(x, y + L)= \psi^a(x, y)\,.
\end{equation}
for simplicity. Then, the same degrees of freedom with the same
poles as the Yang-Mills field can be found in the fermionic field,
except for the replacement $M_A\rightarrow M_{\psi}$.
Thus, we can find that the contribution of the one-loop quantum effects
of fermionic fields becomes
\begin{equation}
V_{\psi}=N_{\psi}\frac{3}{\pi^2L^4}\sum_{\ell=1}^\infty
\frac{w_\psi(M_{\psi}L\ell)}{\ell^5} [2\cos(\ell\phi)+1]\,,
\end{equation}
where $N_{\psi}$ is the number of the fermion fields which belong to the
adjoint representation of $SU(2)$.
The weight factor $w_\psi(x)$ is defined as
\begin{equation}
w_\psi(x)=1-e^{-x}\left(1+x+\frac{1}{3}x^2\right)\,.
\end{equation}
Note that 
$w_\psi(x)\rightarrow 1$ for $x\rightarrow\infty$ and
$w_\psi(x)\rightarrow \frac{1}{6}x^2+O(x^4)$ for $x\rightarrow 0$.

Therefore, the total effective potential for the system under
consideration is given by
\begin{equation}
V_{eff}(\phi)=-\frac{3}{4\pi^2L^4}\sum_{\ell=1}^\infty
\frac{3w_A(M_AL\ell)-4N_{\psi}w_\psi(M_{\psi}L\ell)}{\ell^5}
[2\cos(\ell\phi)+1]\,.
\end{equation}

\begin{figure}[ht]
\centering
\includegraphics[height=5cm]
{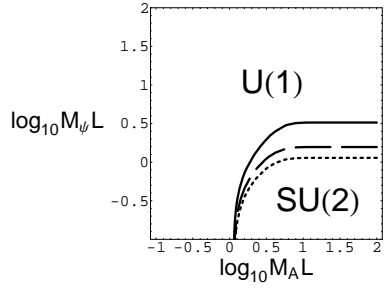}
\caption{%
Phase structure of the model. The solid line indicates the boundary of
the two phases for $N_\psi=1$, the broken line indicates that
for $N_\psi=2$, and the dotted line  indicates that for $N_\psi=3$.}
\label{fig1}
\end{figure}

The emergence of symmetry breaking is indicated by instability of the
symmetric vacuum, or 
\begin{equation}
V''_{eff}(0)=\frac{3}{2\pi^2L^4}\sum_{\ell=1}^\infty
\frac{3w_A(M_AL\ell)-4N_{\psi}w_\psi(M_{\psi}L\ell)}{\ell^3}<0\,.
\end{equation}
The symmetry broken phase is indicated in Fig.~\ref{fig1} in the
parameter space spanned by $M_AL$ and $M_\psi L$.
In the region above the line, the gauge symmetry is broken.
For a sufficiently large $M_\psi L$, the $SU(2)$ symmetry is broken for
any value for $M_A L$.
Conversely speaking, for a sufficiently large $M_A L$, there is a
lower bound value of $M_\psi L$ for symmetry breaking, which is nearly
constant. For $N_\psi=1$, the symmetry breaking occurs if $M_\psi L\ge
3.4$.

\section{Mass of the vector and scalar bosons
\label{S4}}
The phase diagrams are very similar for the different numbers of the
fermions, $N_\psi (\ge 1)$. Thus, from here, we consider only the case of
$N_\psi=1$ in our model.

\begin{figure}[ht]
\centering
\includegraphics[height=5cm]
{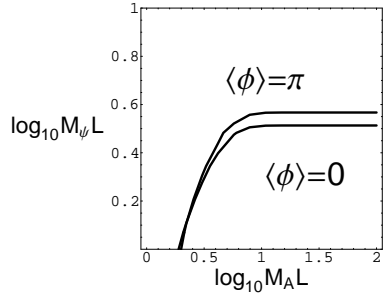}
\caption{%
The value of the order parameter $\langle\phi\rangle$. In the narrow
parameter region between two solid lines, $\langle\phi\rangle$ takes
$0<\langle\phi\rangle<\pi$.}
\label{fig2}
\end{figure}

In our toy model, the order parameter takes the `topological'
value
$\langle\phi\rangle=\pi$ or $\langle\phi\rangle=0$ in the almost region
of the parameter space.
We find, however,  the order parameter $\langle\phi\rangle$ takes the
value $0<\langle\phi\rangle<\pi$ for the narrow region of the parameter
space. This region is shown in Fig.~\ref{fig2} as the narrow region
between two lines.
Because the boundaries are almost parallel to the $M_AL$
axis for a large $M_AL$, we consider the case with
$M_AL\rightarrow\infty$ hereafter.

\begin{figure}[ht]
\centering
\includegraphics[height=5cm]
{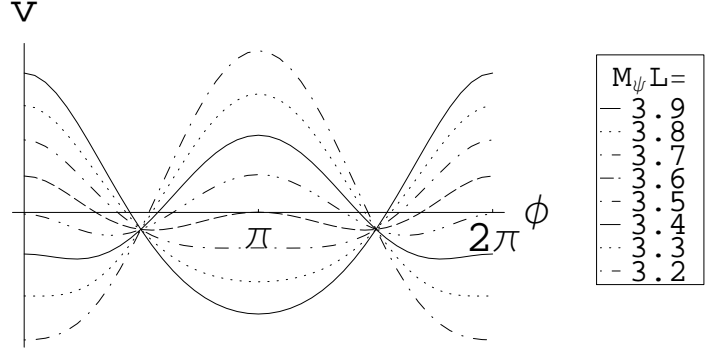}
\caption{%
The profiles of the normalized potential $v(\phi)$ for
$M_AL\rightarrow\infty$ and various values of $M_\psi L$, as indicated in
the legend.}
\label{fig3}
\end{figure}

We define the normalized potential $v(\phi)$ as
$V_{eff}(\phi)=\frac{3}{2\pi^2L^4}v(\phi)+$(independent of $\phi$).
Namely, the normalized potential is written by
\begin{equation}
v(\phi)=-\sum_{\ell=1}^\infty
\frac{3w_A(M_AL\ell)-4w_\psi(M_{\psi}L\ell)}{\ell^5}
\cos\ell\phi\,.
\end{equation}
The shape of the potential $v(\phi)$ (shown in the region $0\le\phi\le
2\pi$) is sensitive against the variation of $M_\psi L$ in the
range from
$3.2$ to
$3.8$, as seen from Fig.~\ref{fig3}. 

As seen from the graph of the potential,
the infinite sum in the expression of the potential can be well
approximated by taking the first two terms in this parameter region.
Thus, the approximate potential is
\begin{eqnarray}
v_{app}(\phi)&=&\left\{1-4e^{-M_\psi L}\left[1+M_\psi L+\frac{(M_\psi
L)^2}{3}\right]\right\}\cos\phi\nonumber \\
& &+\frac{1}{2^5}\left\{1-4e^{-2M_\psi
L}\left[1+2M_\psi L+\frac{4(M_\psi L)^2}{3}\right]\right\}\cos 2\phi\,.
\end{eqnarray}
\begin{figure}[ht]
\centering
\includegraphics[height=5cm]
{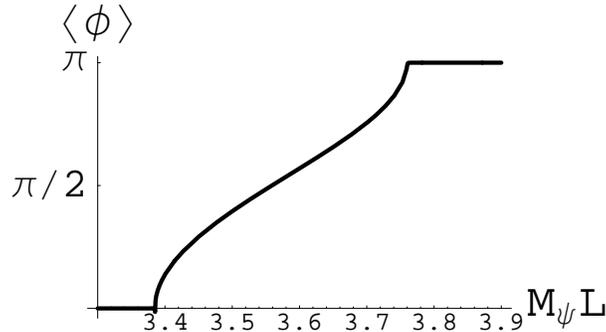}
\caption{%
The vacuum expectation value of $\phi$ against $M_\psi L$.}
\label{fig4}
\end{figure}
Using this approximation, we find the value of the order parameter
$\langle\phi\rangle$ and exhibited in Fig.~\ref{fig4}. We remember that
two vector bosons acquire mass of
$m=\langle\phi\rangle/L$ and one of the gauge bosons remain massless, as
lowest masses of the Kaluza-Klein spectra.
The lowest masses of the adjoint fermions are similar to the vector
bosons.

One light scalar degree of freedom is left in any phases. This is a
scalar field which comes from the extra component of the gauge
field, that is, $\sqrt{L}A_5^{a=3}$.  The scalar is classically
massless, however, as seen so far, the one-loop quantum effect causes
its mass. The mass of this boson is
obtained from the second derivative of the effective potential and we
find
\begin{equation}
m^2_{S}=\frac{3g^2}{2\pi^2L^3}v''(\langle\phi\rangle)\,.
\end{equation}
Note that the dimensionless four dimensional gauge coupling is
$g/\sqrt{L}$ in our model.

\begin{figure}[ht]
\centering
\includegraphics[height=5cm]
{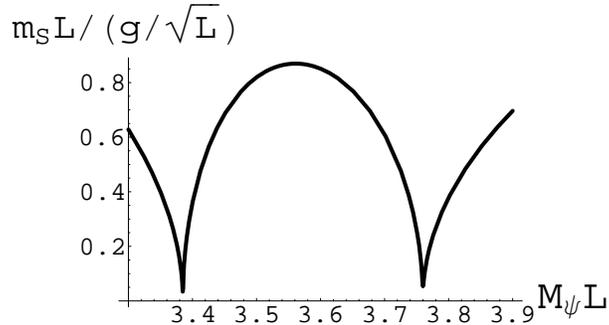}
\caption{%
Mass of the scalar boson.}
\label{fig5}
\end{figure}

The mass of the scalar boson plotted against $M_\psi L$ is
shown in Fig.~\ref{fig5}.
We find that the mass of the scalar boson is always smaller than the mass
of the vector bosons in the symmetry-broken phase, as long as the
four dimensional gauge coupling takes a moderate value $(g/\sqrt{L}<1)$.

\section{Discussion
\label{S5}}
In this paper, we have calculated the one-loop effective potential for
the simple model of five dimensional Lee-Wick gauge theory with adjoint
fermions and have investigated its phase space of the symmetry. It
turns out that the mass of the Lee-Wick particle has much influence on
the symmetry breaking through the effective potential.
We have also found that the smaller mass scales than the
compactification scale
$1/L$ can appear if the Lee-Wick fermion mass scale is in a certain
narrow parameter region.

To study a realistic gauge-Higgs unification scenario, we should
incorporate non-trivial geometry such as orbifolds and
Randall-Sundrum type warped space.
Our toy model, however, has shown the effect of masses of Lee-Wick
particles qualitatively.

In future work, we should examine more elaborated models mentioned above,
their finite temperature behavior (which is interesting as four
dimensional models in Ref.~\cite{thermo}), and higher-loop quantum effect
and non-perturbative effect\footnote{A non-perturbative aspect of the
Hosotani model has been studied by authors of Ref.~\cite{FKP}.} on
the models.  We also wish to study the flux in the extra space in the
Lee-Wick model. In the related thoughts, the possible modification of
Nielsen-Olesen instability
\cite{NO} in the Lee-Wick Yang-Mills theory will also be studied with
much interest.



\end{document}